\documentstyle[12pt,prl,aps,multicol,psfig,epsf]{revtex}
\begin{document}
\sloppy
\thispagestyle{empty}
\begin{flushleft}
DESY 97-236 \hfill {\tt hep-ph/9712219}\\
PRL-TH-97/029\\
December 1997
\end{flushleft}
\mbox{}
\vspace*{\fill}
\begin{center}
{\LARGE\bf On the charm-squark interpretation of the HERA events }

\vspace{1.5em}

\large
{Anjan S.~Joshipura$^{+,*}$, V.~Ravindran$^\dagger$ and Sudhir K.~Vempati$^+$}
\\
\vspace{1.5em}
{\it $^{*}$ Dept. de Fisica Teorica,Univ. of
Valencia,\\461000,Burajssot, Valencia, Spain.
}\\
\vspace{1.5em}
{\it $^\dagger$~DESY--Zeuthen, Platanenallee 6,
D--15735 Zeuthen, Germany}\\
\vspace{2em}
{\it $^+$
Theoretical Physics Group, Physical Research Laboratory,
   \\Navarangpura, Ahmedabad, 380 009, India.}
\end{center}
\vspace*{\fill}
%------------------------------------------------------------
\def\ap#1#2#3{           {\it Ann. Phys. (NY) }{\bf #1} (19#2) #3}
\def\arnps#1#2#3{        {\it Ann. Rev. Nucl. Part. Sci. }{\bf #1} (19#2) #3}
\def\cnpp#1#2#3{        {\it Comm. Nucl. Part. Phys. }{\bf #1} (19#2) #3}
\def\apj#1#2#3{          {\it Astrophys. J. }{\bf #1} (19#2) #3}
\def\asr#1#2#3{          {\it Astrophys. Space Rev. }{\bf #1} (19#2) #3}
\def\ass#1#2#3{          {\it Astrophys. Space Sci. }{\bf #1} (19#2) #3}

\def\apjl#1#2#3{         {\it Astrophys. J. Lett. }{\bf #1} (19#2) #3}
\def\ass#1#2#3{          {\it Astrophys. Space Sci. }{\bf #1} (19#2) #3}
\def\jel#1#2#3{         {\it Journal Europhys. Lett. }{\bf #1} (19#2) #3}

\def\ib#1#2#3{           {\it ibid. }{\bf #1} (19#2) #3}
\def\nat#1#2#3{          {\it Nature }{\bf #1} (19#2) #3}
\def\nps#1#2#3{          {\it Nucl. Phys. B (Proc. Suppl.) }
                         {\bf #1} (19#2) #3} 
\def\np#1#2#3{           {\it Nucl. Phys. }{\bf #1} (19#2) #3}
\def\pl#1#2#3{           {\it Phys. Lett. }{\bf #1} (19#2) #3}
\def\pr#1#2#3{           {\it Phys. Rev. }{\bf #1} (19#2) #3}
\def\prep#1#2#3{         {\it Phys. Rep. }{\bf #1} (19#2) #3}
\def\prl#1#2#3{          {\it Phys. Rev. Lett. }{\bf #1} (19#2) #3}
\def\pw#1#2#3{          {\it Particle World }{\bf #1} (19#2) #3}
\def\ptp#1#2#3{          {\it Prog. Theor. Phys. }{\bf #1} (19#2) #3}
\def\jppnp#1#2#3{         {\it J. Prog. Part. Nucl. Phys. }{\bf #1} (19#2) #3}

\def\rpp#1#2#3{         {\it Rep. on Prog. in Phys. }{\bf #1} (19#2) #3}
\def\ptps#1#2#3{         {\it Prog. Theor. Phys. Suppl. }{\bf #1} (19#2) #3}
\def\rmp#1#2#3{          {\it Rev. Mod. Phys. }{\bf #1} (19#2) #3}
\def\zp#1#2#3{           {\it Zeit. fur Physik }{\bf #1} (19#2) #3}
\def\fp#1#2#3{           {\it Fortschr. Phys. }{\bf #1} (19#2) #3}
\def\Zp#1#2#3{           {\it Z. Physik }{\bf #1} (19#2) #3}
\def\Sci#1#2#3{          {\it Science }{\bf #1} (19#2) #3}
\def\n.c.#1#2#3{         {\it Nuovo Cim. }{\bf #1} (19#2) #3}
\def\r.n.c.#1#2#3{       {\it Riv. del Nuovo Cim. }{\bf #1} (19#2) #3}
\def\sjnp#1#2#3{         {\it Sov. J. Nucl. Phys. }{\bf #1} (19#2) #3}
\def\yf#1#2#3{           {\it Yad. Fiz. }{\bf #1} (19#2) #3}
\def\zetf#1#2#3{         {\it Z. Eksp. Teor. Fiz. }{\bf #1} (19#2) #3}
\def\zetfpr#1#2#3{         {\it Z. Eksp. Teor. Fiz. Pisma. Red. }{\bf #1} (19#2) #3}
\def\jetp#1#2#3{         {\it JETP }{\bf #1} (19#2) #3}
\def\mpl#1#2#3{          {\it Mod. Phys. Lett. }{\bf #1} (19#2) #3}
\def\ufn#1#2#3{          {\it Usp. Fiz. Naut. }{\bf #1} (19#2) #3}
\def\sp#1#2#3{           {\it Sov. Phys.-Usp.}{\bf #1} (19#2) #3}
\def\ppnp#1#2#3{           {\it Prog. Part. Nucl. Phys. }{\bf #1} (19#2) #3}
\def\cnpp#1#2#3{           {\it Comm. Nucl. Part. Phys. }{\bf #1} (19#2) #3}
\def\ijmp#1#2#3{           {\it Int. J. Mod. Phys. }{\bf #1} (19#2) #3}
\def\ic#1#2#3{           {\it Investigaci\'on y Ciencia }{\bf #1} (19#2) #3}
\def\tp{these proceedings}
\def\pc{private communication}
\def\ip{in preparation}
\relax
\newcommand{\GeV}{\,{\rm GeV}}
\newcommand{\MeV}{\,{\rm MeV}}
\newcommand{\keV}{\,{\rm keV}}
\newcommand{\eV}{\,{\rm eV}}
\newcommand{\Tr}{{\rm Tr}\!}
\renewcommand{\arraystretch}{1.2}
\newcommand{\beq}{\begin{equation}}
\newcommand{\eeq}{\end{equation}}
\newcommand{\beqa}{\begin{eqnarray}}
\newcommand{\eeqa}{\end{eqnarray}}
\newcommand{\ba}{\begin{array}}
\newcommand{\ea}{\end{array}}
\newcommand{\bmat}{\left(\ba}
\newcommand{\emat}{\ea\right)}
\newcommand{\refs}[1]{(\ref{#1})}
\newcommand{\ler}{\stackrel{\scriptstyle <}{\scriptstyle\sim}}
\newcommand{\ger}{\stackrel{\scriptstyle >}{\scriptstyle\sim}}
\newcommand{\lag}{\langle}
\newcommand{\rag}{\rangle}
\newcommand{\ns}{\normalsize}
\newcommand{\cm}{{\cal M}}
\newcommand{\gr}{m_{3/2}}
\newcommand{\p}{\partial}

\def\rp{ $R_P$} 
\def\321{$SU(3)\times SU(2)\times U(1)$}
\def\tl{{\tilde{l}}}
\def\tL{{\tilde{L}}}
\def\bd{{\overline{d}}}
\def\tL{{\tilde{L}}}
\def\a{\alpha}
\def\b{\beta}
\def\g{\gamma}
\def\c{\chi}
\def\d{\delta}
\def\D{\Delta}
\def\db{{\overline{\delta}}}
\def\Db{{\overline{\Delta}}}
\def\e{\epsilon}
\def\l{\lambda}
\def\n{\nu}
\def\m{\mu}
\def\nt{{\tilde{\nu}}}
\def\p{\phi}
\def\P{\Phi}
\def\x{\xi}
\def\r{\rho}
\def\s{\sigma}
\def\t{\tau}
\def\th{\theta}
\def\ne{\nu_e}
\def\nm{\nu_{\mu}}
\def\rp{$R_P$}
\def\mp{$M_P$}     
\renewcommand{\Huge}{\Large}
\renewcommand{\LARGE}{\Large}
\renewcommand{\Large}{\large}
%\begin{abstract}
\noindent The charm squark resonance in the $e^+ q$ channel provides one of the
plausible interpretations of the reported anomaly at HERA.
We show that the relevant $R$ violating coupling $\lambda'_{121}$
is required to be large, typically around
  0.1 in a large class of supergravity based models including the
minimal one with the universal boundary condition at the GUT scale.
Existing constraints on these couplings are reanalysed in this light a
nd it is
 argued that such large couplings may be feasible but would require
 fine tuned cancelations.
%\end{abstract}
\newpage
The anomalous events seen by the ZEUS and the H1 detectors \cite{hera}
at HERA in the deep
inelastic $e^+p$ scattering have generated considerable excitement
 \cite{cont,s1,s2,dp1,lepto,kz,others}. These
events would constitute evidence for  physics
beyond standard electroweak model if they are established firmly
in the future. The presently available information when taken seriously
allows two possible interpretations: ({\em i}) The presence of some
lepton number violating contact interaction \cite{cont} or ({\em ii}) 
production of a resonance in the $e^+q$ channel \cite{s1,s2,dp1,lepto}. 
Supersymmetry, 
with violation of $R$ parity\cite{s1,s2,dp1}  provides a
natural theoretical framework to incorporate the second
possibility
although an alternative in terms of a scalar leptoquark \cite{lepto} is 
also open.

The supersymmetric interpretation of the HERA events assumes that the excess
 events seen at
HERA are due to resonant production and  subsequent decay of the squark to $e^+ q$. Three
possibilities have been considered in this context\cite{s1,s2,dp1}:
$e^+_Rd_R\rightarrow \tilde{c}_L,e^+_Rd_R\rightarrow
\tilde{t}_L,e^+_Rs_R\rightarrow \tilde{t}_L$. 
In  analyzing these scenarios \cite{s1,s2,dp1} it has been implicitly
 assumed that the squark masses are free parameters of the
model. While this would be true in the most general situation, 
specific model dependence can alter some of the conclusions. Our
aim is to show that the very minimal model dependent assumption on the
charm squark mass necessarily requires large $\lambda'_{121}$ to
understand HERA events and this large coupling by itself is ruled
out from other constraints. 

The specific assumption that we make and which leads to the above
conclusion is that the charm 
squark mass squared is positive at the unification scale. 
This assumption is  true in the radiative electro-weak breaking
 scenario with 
universal boundary conditions at the GUT scale, but it can also be true in
a much more
general context.  We shall  first assume that the gaugino masses
are unified at $M_{GUT}$ but demonstrate later that 
the removal of this assumption does not
significantly change the basic conclusion.

The argument leading to the above conclusion is largely insensitive to the details
of the radiative $SU(2)\times U(1)$ breaking in  the MSSM and
runs as follows.

Consider the following $R$ violating couplings:
\begin{equation}
\label{wr1}
W_{R} = \lambda'_{ijk} (-\nu_id_{l}  K_{lj}+e_i u_j )d_k^c
\end{equation}
The above terms are defined in the quark mass basis and $K$ denotes the
 Kobayashi-Maskawa 
matrix. The charm squark interpretation of the HERA anomaly requires 
$\lambda'_{121}$ to be non-zero. 
The HERA data can be explained provided
\begin{equation}
\lambda'_{121}\sim \frac{0.025-0.034}{B^{1/2} }
\eeq
The number in the numerator of eq.(2) is indicative  of the required range and
depends upon the weightage given to the different experiments as well as on 
the next to leading order QCD
corrections \cite{kz}.  In the following, we shall take \cite{f1}  the value
  0.025 for
the numerator in the RHS of eq.(2). $B$ refers to the branching ratio for 
the squark decay to $q
e^+$. This decay would take place through the coupling in eq.(1) itself. 
$B$ is also influenced by the $R$ conserving decays of the charm squark to
an $s$ ($c$)  and a chargino  (neutralino) . The $\lambda'_{121}$ and 
the  parameters $\mu,M_2,\tan\beta$ determine $B$ in the MSSM.
HERA data can be reconciled if for a region in these parameters
({\em i}) eq.(2) is satisfied, ({\em ii}) $\lambda'_{121}$ is
consistent with other constraints \cite{apv,kpipi,numass,dbdecay} 
due to $R$ breaking and ({\em iii})
charm squark has a mass around 180- 220 GeV. 

In supergravity based models, the charm squark mass at the weak scale is 
governed by the gauge couplings and 
the gaugino masses. Its value at $Q_0=200 \GeV$ is given in the limit of
neglecting Kobayashi-Maskawa  and $\tilde{c}_L-\tilde{c}_R$ mixing by
\cite{f2} 
\beq
m_{\tilde{c}_L}^2 (Q_0)\approx m_{\tilde{c}_L}^2 (M_{GUT})+
 8.83  M_2^2+ 1/2~ M_Z^2 ~\cos 2\beta~ (1-4/3 \sin^2\theta_W)
\eeq
where we have assumed that the gauge couplings and the gaugino masses are 
unified at the GUT scale, $M_{GUT}=3\times 10^{16}\GeV$ and chosen
 $\alpha_s(M_Z)=0.12$. 
The $M_2$ in eq.(3) is the value of the wino
mass at the weak scale
identified here with $M_Z$. The last term in
the above equation is a (-ve ) contribution from the D-term.
It follows that  the charm squark mass provides  strong upper bound on
$M_2$ as long as $m_{\tilde{c}_L}^2 (M_{GUT})>0$:
\beq
M_2\leq 74.04 \GeV \left({m_{\tilde{c_L}}\over 220 \GeV}\right )
\left ( 1- 0.06 \cos 2\beta \left({220 \GeV \over m_{\tilde{c_L}}}\right
)^2
\right)^{1/2}
\eeq

The branching ratio $B$ is determined in the MSSM by the following widths \cite{kk}:
\beqa
\Gamma (\tilde{c}_L\rightarrow e^+ c)&=& 
{\lambda'^2_{121}\over 16 \pi}
m_{\tilde{c}_L}  \\ 
\Gamma (\tilde{c}_L\rightarrow \chi^0_i c)&=& \frac{\alpha}{2~m_{\tilde{c}_L}^3}
\lambda^{\frac{1}{2}} (m_{\tilde{c}_L}^2, m_{c}^2, m_{\chi^0_i}^2)\nonumber \\
&& \left[~( \mid F_L \mid^2 + \mid F_R \mid^2 ) (m_{\tilde{c}_L}^2 - m_{c}^2 - 
m_{\chi^0_i}^2 ) - 4 m_c m_{\chi ^0_i} Re(F_R 
F_L^\star)~\right] \\ \nonumber
\Gamma (\tilde{c}_L\rightarrow \chi^+_i s)&=& \frac{\alpha}{4 sin^2 \theta_W
m_{\tilde{c}_L}^3} \lambda^{\frac{1}{2}}(m_{\tilde{c}_L}^2, m_{s}^2, 
m_{\chi^+_i}^2) \nonumber \\
&& \left[~( \mid G_L \mid^2 + \mid G_R \mid^2 ) 
(m_{\tilde{c}_L}^2 - m_{s}^2 - m_{\chi^+_i}^ 2 ) -
 4 m_s m_{\chi ^+_i} Re(G_R G_L^\star)~\right] \\ \nonumber
{\mbox where},
F_L&=& \frac{m_c N'^\star_{i4}}{2~ m_W~ sin \theta_W ~sin \beta}, \\ \nonumber
F_R&=& e_c N'_{i1} + \frac{\frac{1}{2} - e_c sin^2 \theta_W}{ cos
\theta_W~ sin \theta_W } N'_{i2}, \\ \nonumber
G_L&=& - \frac{m_s U^\star_{k2} }{\sqrt{2}~ m_W~ cos \beta},\\ \nonumber
G_R&=& V_{k1}.
\eeqa
 We have adopted the same notation as in \cite{kk}.
From the expression for B in terms of the above decay widths, and the HERA
constraint, eq.(2), one can solve for the allowed $\lambda'_{121}$.
The contours in the $\mu-M_2$
plane for different values of $\lambda'_{121}$ are displayed in fig 1a
($\tan\beta$=1) and fig 1b ($\tan\beta=40$). The horizontal lines in these
figures show the upper bound on $M_2$, eq.(4). We also display, the curves
corresponding to two representative values of the chargino masses namely
45 and 85 GeV. The later is the present experimental bound obtained
assuming $R$ conservation. This need not hold in the presence of $R$
violation. It is seen from fig.1b that for chargino mass around 85 GeV,
the bound on $M_2$ by itself rules out charm squark interpretation
for large tan $\beta$ independent of the value of $\lambda'_{121}$ \cite{f3}.
But irrespective of the value of $\tan\beta$ and the chargino mass one
needs very large $\lambda'_{121} \geq 0.13 $ in order to satisfy the
bound on $M_2$ coming
from the charm squark mass. This strong bound on $\lambda'_{121}$
 arises because of
the following reason. For $M_2\leq 74 \GeV$, at least one of the charginos
is sufficiently light and contributes dominantly to the $\tilde{c}_L$
decay. This reduces $B$ \cite{dpf} and results in large value for $\lambda'_{121}$
due to eq.(2). In contrast, the chargino decay is suppressed kinematically
for $\tan \beta\sim 1$ if $M_2> 200\GeV$. This results in smaller allowed value
 as seen from the
figure. But these are in conflict with the charm squark mass.

Let us now see if one could make large $\lambda'_{121}$ consistent with
other constraints. The strong constraints come from  atomic parity violation
 \cite{apv}, the decay $K^+\rightarrow
\pi \nu \bar{\nu}$ \cite{kpipi} and the electron neutrino mass \cite{numass}.  
The recent data from Cs on the relevant weak charge have been argued 
\cite{s2,dp2} to
imply
\beq
\lambda'_{121}\leq 0.074
\eeq
at 2$\sigma$ level in conflict with the large value required here.
In principle, the extra contribution due to
charm squark to atomic parity violation can be canceled by a similar
contribution from the scalar bottom or strange squark but the existing
constraints on the relevant couplings make this cancelation 
 difficult \cite{dp2}.
Thus, one cannot easily avoid the atomic parity violation constraint strictly
in the MSSM but this can be done by postulating new physics, e.g. the
presence of an extra $Z$ \cite{dp2}.

The other significant constraint comes from the decay $K^+\rightarrow \pi
\nu \bar{\nu}$ which implies \cite{kpipi}
\beq
\lambda'_{121}\leq 0.02 \left ({m_{\tilde{d}_R} \over 200 \GeV }\right ).
\eeq
The electron neutrino mass also gives similar constraint in the same
parameter range \cite{numass}. The question of choice of the basis becomes
 relevant in
the discussion of these constraints. This is particularly so when one assumes 
only one
$\lambda'_{ijk}$ to be non-zero. For $\lambda'_{121}$ defined in the mass
basis as in eq.(1) the above constraint is unavoidable if rest of the
couplings are zero. This basis choice is
natural from the point of view of interpreting HERA results but is not
unique. One may redefine the couplings as 
$$ \bar{\lambda'}_{ijk}\equiv K_{jl}\lambda'_{ilk} $$
and rewrite eq.(1) as follows:
\begin{equation}
\label{wr2}
W_{R} =  \bar{\lambda'}_{ijk}(-\nu_id_{j}+e_i K^\dagger_{lj}u_l )d_k^c
\end{equation}
HERA result would now require $\bar{\lambda'}_{121}$ to be large. If this is
the only non-zero $\bar{\lambda'}_{ijk}$ then there will not be any
constraint on $\bar{\lambda'}_{121}$ from the neutrino mass or from 
the $K^+\rightarrow \pi
\nu \bar{\nu}$ decay \cite{kpipi}. But eq.(10) will now generate a contribution to 
the neutrinoless double beta decay which is also severely constrained.
Specifically, one has \cite{dbdecay}
\beq
 K^{\dagger}_{12} \bar{\lambda'}_{121} \leq 2.2 \times 10^{-3}
\left({m_{\tilde{u}_L}\over 200
 \GeV}\right)^2 \left({ m_{\tilde{g}}\over 200
 \GeV}\right)^{1/2} \eeq
This clearly does not allow $\bar{\lambda'}_{121}$ of O(0.1). Thus, 
notwithstanding basis dependence one has problem in accommodating large
value for the relevant coupling. An alternative is to allow more than one 
non-zero
$\bar{\lambda'}_{ijk}$. It is seen from eq.(10) that 
$\bar{\lambda'}_{1j1}$ (j=1,2,3) 
contribute to the neutrinoless $\b\b$ decay and simultaneous presence of
these may lead to cancelations. Eq.(11) now gets replaced by
\beq ( \bar{\lambda'}_{111}+ K^{\dagger}_{12} \bar{\lambda'}_{121}+
K^{\dagger}_{13}\bar{\lambda'}_{131} ) \leq 2.2 \times 10^{-3} 
\left({m_{\tilde{u}_L}\over 200
 \GeV}\right)^2 \left({m_{\tilde{g}}\over 200
 \GeV}\right)^{1/2} \eeq
With $\bar{\lambda'}_{121} \sim 0.13$, cancelation between the  last two terms
is unlikely as it requires $\bar{\lambda'}_{131} \sim 2$. The first two 
terms can cancel but the $\bar{\lambda'}_{111} $ is independently
constrained from the neutrino mass \cite{numass}. Its presence generates a large
contribution to the electron neutrino mass induced through 
neutrino-gaugino mixing \cite{numass}. This  is  given by
\beq
m_{\nu}\sim {g^2\over m_{SUSY}} <\tilde{\nu}>^2
\eeq
The value of the induced sneutrino {\it vev} is sensitive to the MSSM parameters
but can be approximately written as \cite{numass}
\beq
<\tilde{\nu}> \sim {9~\bar{\lambda}_{111}\over 16~
\pi^2}~m_d~~ln \left( \frac{M_{GUT}^2}{M_Z^2} \right)
\eeq
Requiring $m_{\nu}\leq 2 eV$ leads for $m_{SUSY}\sim 100 \GeV$ to 
$$ \bar{\lambda}_{111}\leq .04$$
It is seen that cancelations between the first two terms in
eq.(12) are feasible
and can allow $\bar{\lambda}_{121}\sim 0.13$ if this fine tuning is accepted.
It must be added that the bound in the previous equation is quite
sensitive to the MSSM parameters and for a large range in these
parameters, the actual bound can be stronger \cite{numass} than the
generic bound displayed above.

While  wino and zino control the
decay of the charm squark, its mass is mainly controlled by the large
radiative corrections driven by the gluino mass. The unification of the
gaugino mass parameters relates the two and leads to the above
difficulty. Thus  giving up this unification may open up a
possibility of reconciling 
HERA events . Let us treat   
the gaugino masses $M_{1,2,3}$ at $M_Z$ 
 as independent parameters. Then integration of the RG equation for the
 charm-squark from $M_{GUT}$ to $Q_0=200 \GeV$ leads to  
\beqa
m_{\tilde{c}_L}^2 (Q_0)&\approx& m_{\tilde{c}_L}^2 (M_{GUT})+
 0.77  M_3^2 +0.70 M_2^2 + 0.024 M_1^2 \nonumber \\
&+&1/2~ M_Z^2~ \cos 2\beta~ (1-4/3 \sin^2\theta_W)
\eeqa
If gaugino masses were to be unified at $M_{GUT}$ then $M_3\sim 3.25 M_2$
and $M_1\sim 0.5 M_2$. Even in the absence of such unification, the
physical gluino mass $m_{\tilde g}\sim (1+4.2 {\alpha_s\over \pi}
)M_3$ must be greater than
the charm squark mass if large $\lambda'_{121}$ is to be avoided.
This follows since in the converse case, the charm squark would
predominantly decay to a gluino and a quark. This decay being governed
by strong coupling, would dominate the other decays and would reduce
$B$. The later is given in case of $m_{\tilde{c}_L}\gg
m_{\tilde{g}}$  by
$$
B\sim {3~\lambda'^2_{121}\over 32~ \pi~\alpha_s} \sim 2.5 \times
\;10^{-3}
\left({\lambda'_{121}\over 0.1}\right)^2$$
Such a tiny value of $B$ would need unacceptably large $\lambda'_{121}$.
It therefore follows that  one must suppress the squark decay to gluino
 kinematically by
requiring $m_{\tilde {g}}\geq m_{\tilde {c}_L}$. Given this bound on
 $M_3$ it follows from
eq.(15) that 
\beq
M_2\leq 170 \GeV
\eeq
if $m_{\tilde{c}_L}\sim 220 \GeV$. This bound on $M_2$ is  weaker than
 the one in the case of the 
gaugino mass unification, eq.(4). But it nevertheless cannot suppress
the 
decay of squarks to chargino kinematically. It follows
\cite{f4} from Fig. 1  that one now  approximately needs 
$ \lambda'_{121}\geq 0.08$. This value is close to the 
2$\sigma$ limit coming from the atomic parity violation but 
one would still need some cancelations to satisfy other
constraints as discussed above.
Thus giving up
unification helps only partially.

An alternative possibility is to allow for a -ve (mass)$^2$ for the charm
squark at the unification scale. In view of the large positive
contribution induced by the gluino mass such negative (mass)$^2$ need not lead
to colour breaking and may be consistent phenomenologically. In fact  a
 -ve (mass)$^2$ for top squark has been considered
in the literature \cite{negativemass} in a different context.
 The universality is a simplifying feature of 
MSSM but it does not follow from any general principle. It
 does not hold in a large class of string based models
which may allow negative (mass)$^2$ for some sfermions as well
\cite{string}. Such masses  can also arise  when  SUSY is
broken by an anomalous $U(1)$ \cite{u1} with some of the sparticles
 having -ve charge
under this $U(1)$. 

The large radiative corrections induced through the running
in squark masses from a high $\sim M_{GUT}$ to the weak
scale has played an important role in this analysis. In
contrast to the supergravity based models, this
running is over a much smaller range in models with gauge
mediated supersymmetric breaking. But in these models, the
initial value of the charm squark (mass)$^2$ is positive and
large with the result that these models are
incompatible with the charm squark interpretation of HERA 
anomaly even  without the radiative corrections
\cite{f5}.

The interpretation of HERA events in terms of stop may not suffer from the
above mentioned difficulty encountered for the charm interpretation for two
reasons. Firstly, the stop mass is reduced compared to the charm squark
mass due to the  possible large
$\tilde{t}_L-\tilde{t}_R$ mixing as well due to the large top coupling.
Secondly, this mass also
involves one more parameter (the trilinear couplings $A$) compared to the
charm squark mass. Thus while this is a less constrained possibility,
imposition of the requirement that $m_{\tilde{t}}\sim 200 \GeV$ would
certainly lead to more constrained parameter space than considered 
in model independent studies \cite{s1}.

In summary, we have shown that the charm squark interpretation of HERA events
is possible only for large $\lambda'_{121}\sim O( 0.1)$ in a large class of
supersymmetric standard models characterized by a positive charm squark 
(mass)$^2$ at the GUT scale. The simplest and the most popular
minimal supergravity model with universal boundary condition falls in
this class. The required large value of $R$ violating parameter
is difficult to admit without postulating new physics and /or fine
tuned cancelations.

\noindent
{\bf Acknowledgments:}~~ We gratefully acknowledge very helpful comments
from J. Bl\"{u}mlein, Debajyoti Chaudhury, S. Raychaudhuri and D.P. Roy.  

%\newpage

\newpage
\begin{figure}[h]
\epsfxsize 15 cm
\epsfysize 15 cm
\epsfbox[25 151 585 704]{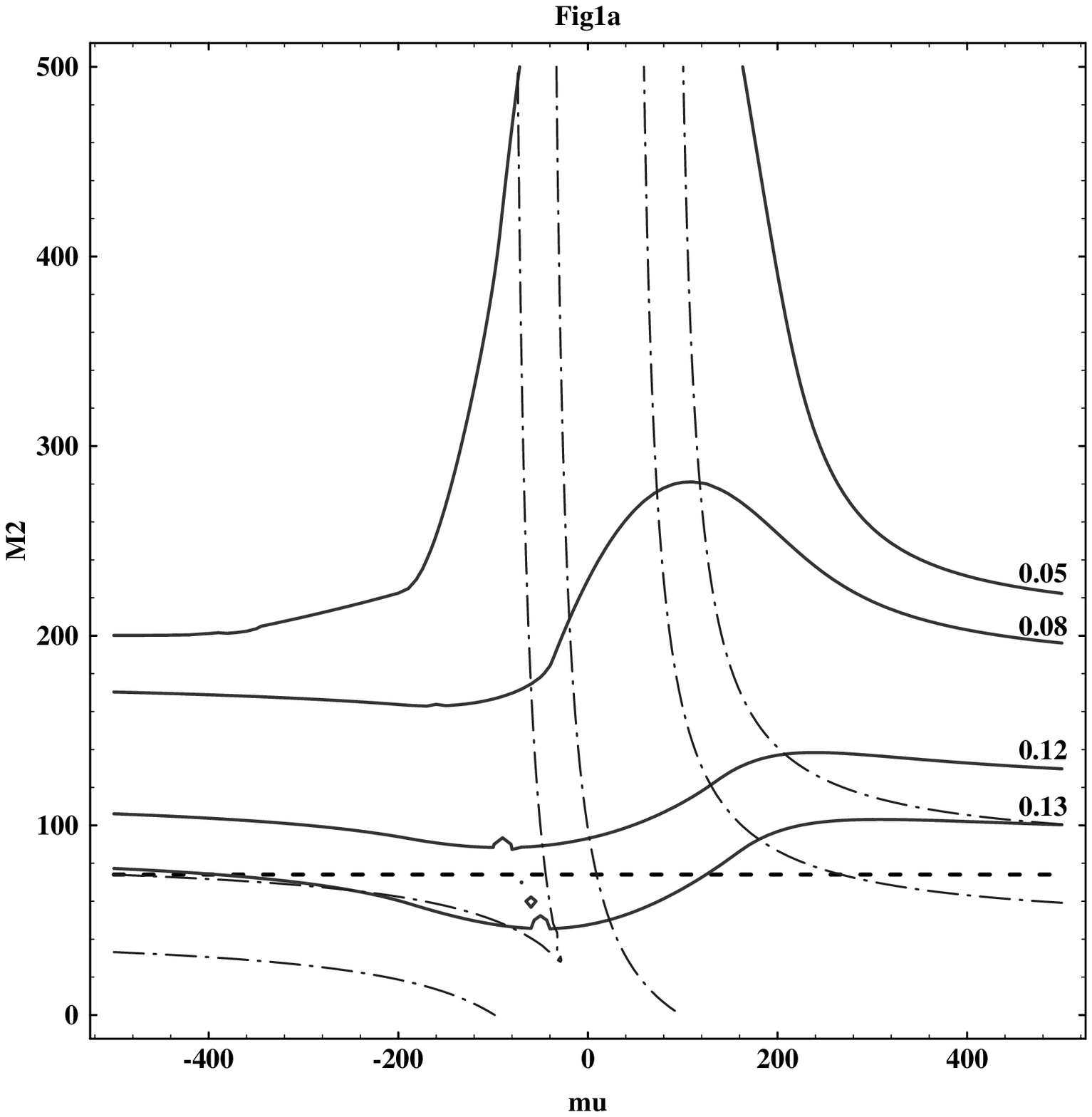}
\label{fig1a}
\end{figure}
\noindent
{\bf Fig 1 a:} ~\sl{The contours (continuous lines ) of 
constant $\lambda'_{121}$  
obtained by imposing HERA constraint, eq.(2). The contours are for
values  0.05, 0.08, 0.12, and 0.13. 
The horizontal dashed line represents the bound on $M_2 $ 
coming from requiring $m_{\tilde{c}_L}=220 \GeV$. 
The vertical dash-dot lines represent the bounds on 
the chargino mass, the upper one for a mass of 85 \GeV and the
lower one for
a mass of 45 \GeV.  All the
above are computed for tan$\beta$ = 1. }

\begin{figure}[h]
\epsfxsize 15 cm
\epsfysize 15 cm
\epsfbox[25 151 585 704]{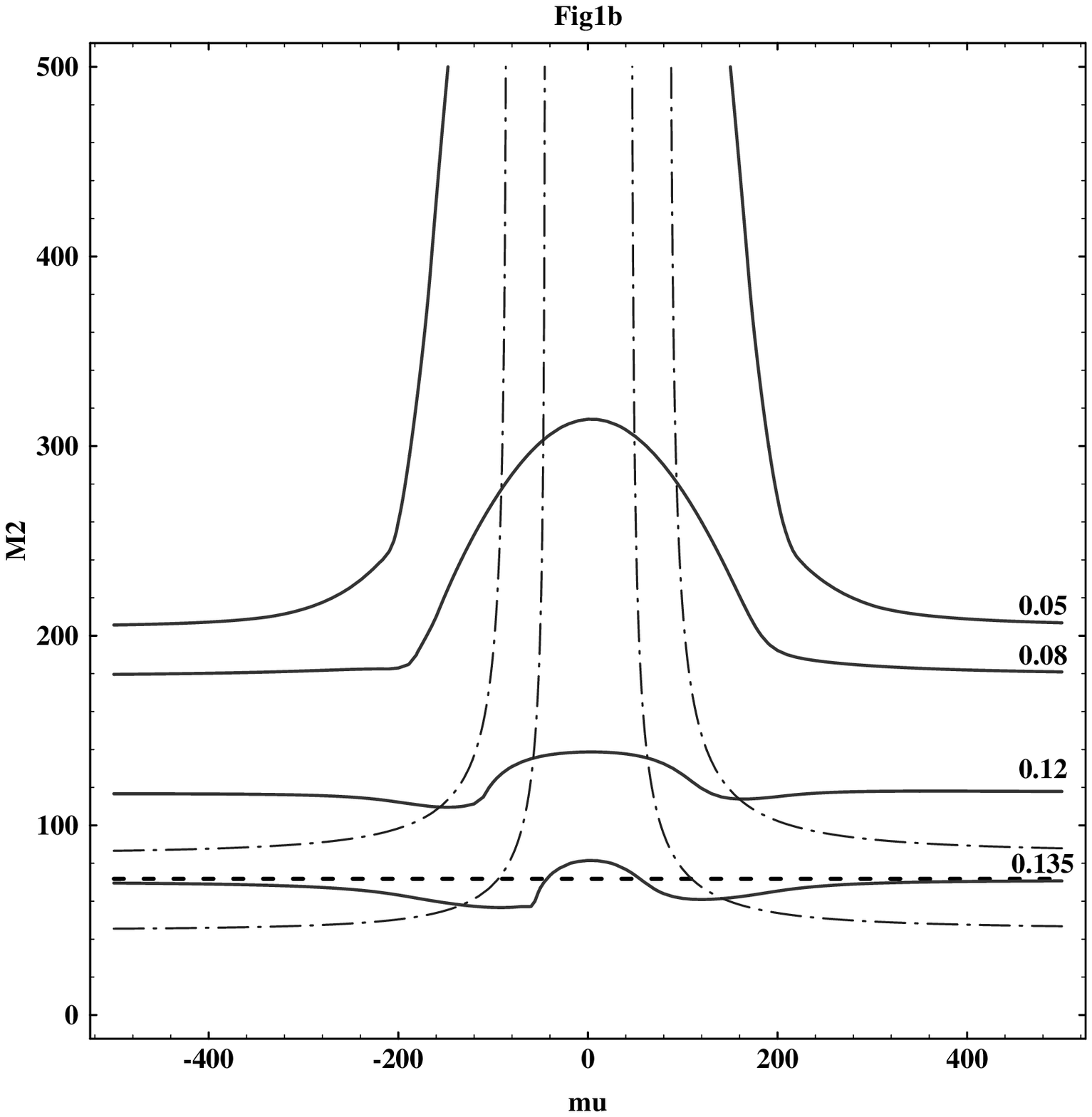}
\label{fig1b}
\end{figure}
\noindent
{\bf Fig 1 b:}~\sl{ Same as fig 1a but  for tan$\beta$ = 40 and
$\lambda'_{121}$=0.05, 0.08, 0.12, and 0.135.}


\begin {thebibliography}{99}
\bibitem{f2} We also neglect the effect of additional trilinear $R$ violating
couplings on the running of the charm squark mass. Their inclusion does
not significantly alter the charm squark mass even when they 
are large, see e.g.  K. Cheung, D. Dicus and B. Dutta \cite{s1}.
\bibitem{f3} Specifically, the upper bound on $M_2$ can be reconciled
with the chargino mass of 85 GeV or more only if $\tan\beta\leq 2.5$.
\bibitem{dpf} Reduction in the branching ratio for charm squark decay
in case of the minimal model was also noticed in \cite{dp1}.
\bibitem{f4} Note that fig. 1 is based on the assumption of $M_1=.5
M_2$ but does not use any relation between  $M_3$ and $M_2$.
\end{thebibliography}
\end{document}